\documentclass[ALICE,manyauthors]{cernphprep}
\usepackage[comma,square,numbers,sort&compress]{natbib}

\usepackage{lineno}
\usepackage{xspace}
\usepackage{hyperref}

\usepackage[inline]{enumitem}
\usepackage[T1]{fontenc}
\usepackage{orcidlink}

\usepackage{color}

\begin{document}
%

\newcommand{\pp}           {pp\xspace}
\newcommand{\ppbar}        {\mbox{$\mathrm {p\overline{p}}$}\xspace}
\newcommand{\XeXe}         {\mbox{Xe--Xe}\xspace}
\newcommand{\PbPb}         {\mbox{Pb--Pb}\xspace}
\newcommand{\pA}           {\mbox{pA}\xspace}
\newcommand{\pPb}          {\mbox{p--Pb}\xspace}
\newcommand{\AuAu}         {\mbox{Au--Au}\xspace}
\newcommand{\dAu}          {\mbox{d--Au}\xspace}

\newcommand{\s}            {\ensuremath{\sqrt{s}}\xspace}
\newcommand{\snn}          {\ensuremath{\sqrt{s_{\mathrm{NN}}}}\xspace}
\newcommand{\pt}           {\ensuremath{p_{\rm T}}\xspace}
\newcommand{\meanpt}       {$\langle p_{\mathrm{T}}\rangle$\xspace}
\newcommand{\ycms}         {\ensuremath{y_{\rm CMS}}\xspace}
\newcommand{\ylab}         {\ensuremath{y_{\rm lab}}\xspace}
\newcommand{\etarange}[1]  {\mbox{$\left | \eta \right |~<~#1$}}
\newcommand{\yrange}[1]    {\mbox{$\left | y \right |~<~#1$}}
\newcommand{\dndy}         {\ensuremath{\mathrm{d}N_\mathrm{ch}/\mathrm{d}y}\xspace}
\newcommand{\dndeta}       {\ensuremath{\mathrm{d}N_\mathrm{ch}/\mathrm{d}\eta}\xspace}
\newcommand{\avdndeta}     {\ensuremath{\langle\dndeta\rangle}\xspace}
\newcommand{\dNdy}         {\ensuremath{\mathrm{d}N_\mathrm{ch}/\mathrm{d}y}\xspace}
\newcommand{\Npart}        {\ensuremath{N_\mathrm{part}}\xspace}
\newcommand{\Ncoll}        {\ensuremath{N_\mathrm{coll}}\xspace}
\newcommand{\dEdx}         {\ensuremath{\textrm{d}E/\textrm{d}x}\xspace}
\newcommand{\RpPb}         {\ensuremath{R_{\rm pPb}}\xspace}

\newcommand{\nineH}        {$\sqrt{s}~=~0.9$~Te\kern-.1emV\xspace}
\newcommand{\seven}        {$\sqrt{s}~=~7$~Te\kern-.1emV\xspace}
\newcommand{\twoH}         {$\sqrt{s}~=~0.2$~Te\kern-.1emV\xspace}
\newcommand{\twosevensix}  {$\sqrt{s}~=~2.76$~Te\kern-.1emV\xspace}
\newcommand{\five}         {$\sqrt{s}~=~5.02$~Te\kern-.1emV\xspace}
\newcommand{\twosevensixnn}{$\sqrt{s_{\mathrm{NN}}}~=~2.76$~Te\kern-.1emV\xspace}
\newcommand{\fivenn}       {$\sqrt{s_{\mathrm{NN}}}~=~5.02$~Te\kern-.1emV\xspace}
\newcommand{\LT}           {L{\'e}vy-Tsallis\xspace}
\newcommand{\GeVc}         {Ge\kern-.1emV/$c$\xspace}
\newcommand{\MeVc}         {Me\kern-.1emV/$c$\xspace}
\newcommand{\TeV}          {Te\kern-.1emV\xspace}
\newcommand{\GeV}          {Ge\kern-.1emV\xspace}
\newcommand{\MeV}          {Me\kern-.1emV\xspace}
\newcommand{\GeVmass}      {Ge\kern-.2emV/$c^2$\xspace}
\newcommand{\MeVmass}      {Me\kern-.2emV/$c^2$\xspace}
\newcommand{\lumi}         {\ensuremath{\mathcal{L}}\xspace}

\newcommand{\ITS}          {\rm{ITS}\xspace}
\newcommand{\TOF}          {\rm{TOF}\xspace}
\newcommand{\ZDC}          {\rm{ZDC}\xspace}
\newcommand{\ZDCs}         {\rm{ZDCs}\xspace}
\newcommand{\ZNA}          {\rm{ZNA}\xspace}
\newcommand{\ZNC}          {\rm{ZNC}\xspace}
\newcommand{\SPD}          {\rm{SPD}\xspace}
\newcommand{\SDD}          {\rm{SDD}\xspace}
\newcommand{\SSD}          {\rm{SSD}\xspace}
\newcommand{\TPC}          {\rm{TPC}\xspace}
\newcommand{\TRD}          {\rm{TRD}\xspace}
\newcommand{\VZERO}        {\rm{V0}\xspace}
\newcommand{\VZEROA}       {\rm{V0A}\xspace}
\newcommand{\VZEROC}       {\rm{V0C}\xspace}
\newcommand{\Vdecay} 	   {\ensuremath{V^{0}}\xspace}

\newcommand{\ee}           {\ensuremath{e^{+}e^{-}}} 
\newcommand{\pip}          {\ensuremath{\pi^{+}}\xspace}
\newcommand{\pim}          {\ensuremath{\pi^{-}}\xspace}
\newcommand{\kap}          {\ensuremath{\rm{K}^{+}}\xspace}
\newcommand{\kam}          {\ensuremath{\rm{K}^{-}}\xspace}
\newcommand{\pbar}         {\ensuremath{\rm\overline{p}}\xspace}
\newcommand{\kzero}        {\ensuremath{{\rm K}^{0}_{\rm{S}}}\xspace}
\newcommand{\lmb}          {\ensuremath{\Lambda}\xspace}
\newcommand{\almb}         {\ensuremath{\overline{\Lambda}}\xspace}
\newcommand{\Om}           {\ensuremath{\Omega^-}\xspace}
\newcommand{\Mo}           {\ensuremath{\overline{\Omega}^+}\xspace}
\newcommand{\X}            {\ensuremath{\Xi^-}\xspace}
\newcommand{\Ix}           {\ensuremath{\overline{\Xi}^+}\xspace}
\newcommand{\Xis}          {\ensuremath{\Xi^{\pm}}\xspace}
\newcommand{\Oms}          {\ensuremath{\Omega^{\pm}}\xspace}
\newcommand{\degree}       {\ensuremath{^{\rm o}}\xspace}

\newcommand{\Jpsi}{\rm{J}/\psi}
\newcommand{\psit}{\psi(2{\rm S})}
\newcommand{\psip}{\psi(2{\rm S})}
\newcommand{\psipOJpsi}{\ensuremath{\psip\mbox{-over-}\Jpsi}}
\newcommand{\Tm}{\ensuremath{\rm{T}\cdot}\rm{m}\xspace}
\newcommand{\abs}[1]{\left\vert #1 \right\vert}
\newcommand{\ave}[1]{\langle #1 \rangle}
\newcommand{\zv}{z_{\rm vtx}}
\newcommand{\Ntr}{N_{\rm tracklet}}
\newcommand{\avNtr}{\langle N_{\rm tracklet} \rangle}
\newcommand{\Ntrc}{N_{\rm tracklet}^{\rm corr}}
\newcommand{\Ncorr}        {\ensuremath{N^\mathrm{corr}}\xspace}
\newcommand{\Ae}{A\varepsilon}
\newcommand{\mumu}{\mu^+\mu^-}
\newcommand{\BR}{\rm{BR}_{\Jpsi\rightarrow\mumu}}

\newcommand{\Ups}[1]{\Upsilon(#1{\rm S})}
\newcommand{\eightnn}       {$\sqrt{s_{\mathrm{NN}}}~=~8.16$~Te\kern-.1emV\xspace}
\newcommand{\AAcoll}          {\mbox{A--A}\xspace}

\newcommand{\thirteen}        {$\sqrt{s}~=~13$~Te\kern-.1emV\xspace}

\newcommand{\seff}{\ensuremath\sigma_{\rm eff}}
\newcommand{\QQbar}{{\rm Q}\overline{\rm Q}}
\newcommand{\ccbar}{{\rm c}\overline{\rm c}}
\newcommand{\bbbar}{{\rm b}\overline{\rm b}}

\begin{titlepage}
\PHyear{2023}       
\PHnumber{053}      
\PHdate{22 March}  

\title{Measurement of inclusive $\Jpsi$ pair production cross section in pp collisions at \thirteen}
\ShortTitle{Inclusive $\Jpsi$ pair production in pp collisions at \thirteen}   

\Collaboration{ALICE Collaboration\thanks{See Appendix~\ref{app:collab} for the list of collaboration members}}
\ShortAuthor{ALICE Collaboration} 

\begin{abstract}
The production cross section of inclusive $\Jpsi$ pairs in \pp collisions at a centre-of-mass energy \thirteen is measured with ALICE. The measurement is performed for $\Jpsi$ in the rapidity interval $2.5<y<4.0$ and for transverse momentum $\pt>0$. 
The production cross section of inclusive $\Jpsi$ pairs is reported to be $10.3 \pm 2.3 \, {\rm (stat.)} \pm 1.3 \, {\rm (syst.)}$~nb in this kinematic interval. 
The contribution from non-prompt $\Jpsi$ (i.e.\ originated from beauty-hadron decays) to the inclusive sample is evaluated. 
The effective double-parton scattering cross section is computed, neglecting the single-parton scattering contribution.

\end{abstract}
\end{titlepage}

\setcounter{page}{2} 


\section{Introduction}
\label{sec:intro}


In the quantum chromodynamics (QCD) parton model~\cite{Bjorken:1969ja}, hadrons are composed of elementary constituents, the partons. Due to the composite nature of hadrons, multiple parton hard scatterings can occur in a hadron--hadron collision. Thus, it is possible to have two or more hard parton interactions simultaneously. 
Multiple parton interactions (MPI) have been studied since the introduction of the parton model~\cite{Takagi:1979wn,Goebel:1979mi}. 
Further studies included the generalization of the QCD evolution equations into multiparton distribution and fragmentation functions 
\cite{Kirschner:1979im,Shelest:1982dg}, 
and a discussion on the possible correlations in the colour and spin degrees of freedom~\cite{Mekhfi:1985dv}. 
Double-parton scatterings (DPS) are the simplest case of MPI, 
and were found to play the most important role in processes with final states such as four jets, four leptons or $n$-jet + W/$\gamma$ measurements 
\cite{Paver:1982yp,Paver:1983hi,Mekhfi:1983az,Humpert:1983pw,Humpert:1983fy,Humpert:1984ay,Ametller:1985tp,Halzen:1986ue,Godbole:1989ti}. 
These studies were complemented by several measurements in hadron collisions at center-of-mass energies ($\s$) ranging from 63 GeV to 1.96 TeV ~\cite{AxialFieldSpectrometer:1986dfj,UA2:1991apc,CDF:1993sbj,CDF:1997lmq,D0:2009apj,D0:2014owy,D0:2014vql}.

At the LHC energies, the probability to have multiple parton interactions increases: as with the increase of collision energy, partons with smaller momentum fraction $x$ are probed with larger fluxes.
Recent measurements have shown the relevance of MPI at the LHC 
\cite{LHCb:2012aiv,ATLAS:2013aph,CMS:2013huw,ATLAS:2014ofp,LHCb:2015wvu,ALICE:2022wpn}, 
and have contributed to stimulate recent progress in the theoretical understanding of MPI~\cite{Blok:2011bu,Diehl:2011yj,Gaunt:2014ska,Diehl:2014vaa,Gaunt:2014rua}. 
Nevertheless, a quantitative estimate of the DPS impact on observables remains challenging. 
Neglecting the parton correlations in the proton, the DPS contribution to a final state \mbox{$A + B$} can be evaluated as the product of the parton level cross sections ($\widehat{\sigma}$) divided by an effective cross section ($\seff$)~\cite{Humpert:1984ay,Ametller:1985tp,DelFabbro:2000ds} 
\begin{equation}
\sigma^{\rm DPS}_{A,B} = \frac{m}{2} \,\frac{\widehat{\sigma}^{A} \, \widehat{\sigma}^{B}}{\seff}   \,,
\label{eq:DPSred} 
\end{equation}
 where the parameter $m$ is a symmetry factor, $m=1$ if $A=B$, and 2 otherwise. The effective cross section is a phenomenological parameter related to the transverse overlap function between the partons of the proton, and is thought to be universal. 
It was found to range between 2 and 25 mb 
\cite{D0:2014vql,ATLAS:2016ydt,LHCb:2016wuo,D0:2015dyx,Lansberg:2016muq,Lansberg:2017chq,LHCb:2015wvu,LHCb:2012aiv,CDF:1997yfa,D0:2009apj,CDF:1993sbj,CMS:2015wcf,CMS:2013huw,ATLAS:2013aph,CMS:2019jcb}.

Double particle production is typically exploited to study DPS. 
A non exhaustive list of these studies are the measurements of the production cross sections of double quarkonium,  i.e.\ ${\rm J}/\psi$ pairs~\cite{NA3:1982qlq,PLB198585,ATLAS:2016ydt,CMS:2014cmt,LHCb:2011kri,LHCb:2016wuo,D0:2014vql}, $\Upsilon$ pairs~\cite{CMS:2016liw}, or ${\rm J}/\psi + \Upsilon$~\cite{D0:2015dyx}, 
electroweak boson plus quarkonium 
\cite{D0:2014vql,ATLAS:2014jkm,ATLAS:2014ofp,Lansberg:2016muq,Lansberg:2017chq,LHCb:2015wvu}, 
double charm production~\cite{LHCb:2012aiv},
charmed hadrons plus quarkonium~\cite{LHCb:2015wvu,LHCb:2012aiv}, 
electroweak boson plus open charm ~\cite{LHCb:2012aiv,LHCb:2014kke}, 
 as well as measurements with jets in the final state, multi-jets~\cite{AxialFieldSpectrometer:1986dfj,UA2:1991apc,CDF:1993sbj,ATLAS:2016rnd}, 
$\gamma + 3$-jets~\cite{CDF:1997lmq,D0:2009apj,D0:2014owy}, 
$2\gamma + 2$-jets~\cite{D0:2015rpo}, and
W + 2-jets~\cite{CMS:2013huw}. 
The recent observation of triple J/$\psi$ production proposes an additional channel to study double and triple parton scatterings 
\cite{CMS:2021qsn}.

 In the quarkonium sector, quarkonium-pair production is a golden tool to probe the production mechanism of heavy quarkonia~\cite{Ko:2010xy,Sun:2014gca,Baranov:2015cle}. 
 The production mechanism of heavy quarkonia is not fully understood after more than forty years of study, and considered a long-standing puzzle of QCD. 
The colour-singlet model (CSM), which assumes the formation of an intermediate $\QQbar$ state with the quantum numbers of the final state, underestimates the production cross section at high $\pt$ both at leading-order (LO) and next-to-leading-order (NLO)~\cite{Lansberg:2011hi,Campbell:2007ws,Gong:2008sn}. 
The recent CSM 
next-to-next-to-leading-order
NNLO$^\star$ calculations have reduced the discrepancies~\cite{Lansberg:2011hi,Lansberg:2019adr}. 
Non-relativistic QCD (NRQCD) calculations consider both colour-singlet (CS) and colour-octet (CO) states of the $\QQbar$ pair~\cite{Bodwin:1994jh}, but fail to predict at the same time the production cross section and polarisation~\cite{LHCb:2013izl,LHCb:2014brf,CMS:2012bpf,CMS:2013gbz,ALICE:2011gej}. 
The selection rules for pair production in the CS process of LO NRQCD forbid the feed-down from cascade decays of excited charge-conjugate-even states, e.g.\ $\chi_{\rm c} \rightarrow {\rm J}/\psi \, \gamma$, whose contribution is significant in single quarkonium production, and makes the comparison of data to model calculations difficult.
As a consequence, quarkonium-pair production provides stringent tests of model calculations.

In this letter we report the measurement of inclusive $\Jpsi$ pair production cross section in \pp collisions at \mbox{$\s=13$~TeV} at large rapidity ($2.5<y<4.0$) with ALICE. 
Inclusive $\Jpsi$ results correspond to the sum two contributions: the prompt contribution, originated from direct charm decays or decays of higher-mass excited states; 
and the non-prompt contribution, stemming from beauty decays.
The results corroborate analogous measurements performed in a similar rapidity interval by LHCb~\cite{LHCb:2016wuo}. 
They constitute a probe to study quarkonium production mechanisms and the DPS contribution.

\section{Experimental apparatus and data sample}
\label{sec:detector}

A description of the ALICE detector and its performances can be found in Refs.~\cite{ALICE:2008ngc,ALICE:2014sbx}.
At forward rapidity ($2.5<y<4.0$) the production of quarkonium states is measured in the muon spectrometer down to $\pt = 0$ via their dimuon decay channel. 
The muon spectrometer of ALICE consists of a ten interaction length thick  front absorber to filter muons, five tracking stations of two planes of cathode pad chambers each (MCH), a dipole magnet with a field integral of 3 Tm surrounding the third tracking station, a 1.2 m thick iron wall to absorb secondary hadrons escaping from the front absorber and low momentum muons coming mainly from $\pi$ and K decays, and two trigger stations made of two planes of resistive plate chambers each (MTR)~\cite{ALICE:2011zqe}. 
The silicon pixel detector (\SPD) and scintillator arrays (\VZERO) are also used in this analysis. The \VZERO counters, two arrays of 32 scintillator tiles each, cover $2.8 \leq \eta \leq 5.1$ (\VZEROA) and $-3.7 \leq \eta  \leq -1.7 $  (\VZEROC) and provide trigger information. The minimum bias (MB) trigger requirement consists of a logical AND of  a signal in \VZEROA and in \VZEROC. The \SPD,  two cylindrical layers covering $|\eta| \leq  2.0$ and $|\eta| \leq 1.4$ for the inner and outer layers, respectively, is dedicated to the  vertex reconstruction and allows estimating pile-up. 
The maximum interaction rate for the analysed data sample is 260 kHz, and the maximum pile-up probability is about $5\times10^{-3}$, negligible for this measurement.

The $\Jpsi$ pair analysis is performed using data from pp collisions at \thirteen collected from 2016 to 2018. 
The event sample was selected using the dimuon trigger condition, which 
is defined as  the  coincidence between the MB requirement and two opposite-charge sign track segments in the muon spectrometer trigger stations. Each track segment in the trigger stations is required to have a transverse momentum, evaluated online, larger than about 0.5~\GeVc. 
Only events passing a selection criterion to remove beam--background collisions contamination, based on the timing information from the V0 arrays, are considered in the analysis. 

When multiple primary vertices are reconstructed by the \SPD, the event is tagged as pile-up and removed from this analysis. 
In order to avoid acceptance biases on the reconstructed \SPD tracklets, events with a displaced vertex with respect to centre of the \SPD detector along the beam direction are discarded according to the requirement $|v_{z}| \leq 10$~cm. 
These selections allowed us to keep the pile-up below $0.3\%$ for the analysed events, also for events with two muon pairs with an invariant mass above 2 \GeVmass. 
Considering the above selections, the total number of dimuon triggered events in the sample sums up to $587.4 \times 10^{6}$ events and corresponds to an integrated luminosity of $24.11 \pm  0.01 (\rm{stat.}) \pm 0.80 (\rm{syst.}) \, \rm{pb}^{-1}$.

\section{Analysis}
\label{sec:analysis}

$\Jpsi$ candidates are built from  muon pairs of opposite-charge sign. 
Muons are identified by requiring that selected tracks in the MCH have a matching track segment in MTR. 
Only muon tracks within the detector acceptance are kept for analysis. Tracks are required to be within $-4.0 < \eta^{\mu} < -2.5$, and the radial distance from the beam axis at the end of the front absorber, $R_{\rm abs}$, is limited to $17.6 < R_{\rm abs} < 89.5$~cm~\cite{ALICE:2017leg}. 
$\Jpsi$ pair candidates are reconstructed from all combinations of double dimuon pairs (each dimuon consisting of an opposite-charge sign muon pair) per event.

The production cross section of inclusive $\Jpsi$ pairs is determined as
\begin{equation}
    \sigma ({\rm J}/\psi \,\, {\rm J}/\psi ) 
     = \frac{N}{ \mathcal{L}_{\rm int} \times \epsilon \times B^2( {\rm J}/\psi \rightarrow \mu^+ \mu^-)} \, ,
\end{equation}
where $N$ is the signal estimate, $\epsilon$ is the acceptance-times-efficiency correction, $B(\Jpsi \rightarrow \mu^+ \mu^-)=(5.961 \pm 0.033) \%$ is the branching fraction of $\Jpsi \rightarrow \mu^+ \mu^-$~\cite{Workman:2022ynf}, and $\mathcal{L}_{\rm int} $ is the integrated luminosity. 
%

\begin{figure}[!htbp]
    \begin{center}
    \includegraphics[width = 0.98\textwidth]{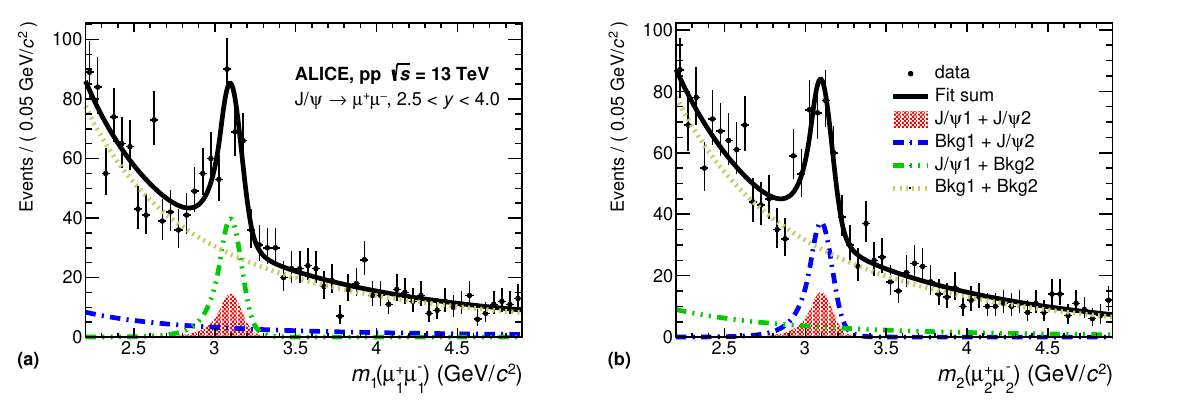}
    \end{center}
    \caption{Projections of a fit to the 2-dimensional invariant mass distribution for inclusive $\Jpsi$ pair candidates for (a) $m_1(\mu_1^+ \mu_1^-)$ and (b) $m_2(\mu_2^+ \mu_2^-)$. 
    The (black) markers show data. 
    The (black) solid line represents the total fit function. 
    The (blue and green) dashed-dotted lines indicate the background contribution from a combination of a real $\Jpsi$ signal with a combinatorial candidate. 
    The (yellow) dotted line represents muon pairs from combinatorial background. 
    \label{fig:invmass}
    }
\end{figure}

%
The $\Jpsi$ pair signal is evaluated from a fit to the 2-dimensional invariant mass distribution. A two-step procedure was chosen. 
The first step exploits the 1-dimensional distribution of all $\Jpsi$ candidates in the data sample analysed, to obtain a good description of the $\Jpsi$ line shape from data. 
A fit is performed with a superposition of $\Jpsi$ and $\psi(2S)$ signal functions and a background function. 
The $\Jpsi$ mass, width, and normalisation are left free in the procedure. 
Instead, the $\psi(2S)$ mass and width are bound to those of $\Jpsi$ as described in Ref.~\cite{ALICE:2015pgg}. 
The 2-dimensional invariant mass distribution of $\Jpsi$ pair candidates ($m_1(\mu_1^+ \mu_1^-)$, $m_2(\mu_2^+ \mu_2^-)$) is fit in the second step via $F(m_1,m_2)$. 
\begin{eqnarray}
F(m_1,m_2) & = & N \times S_1 (m_1) \times S_2 (m_2) 
        + R_{B_1, S_2} \times B_1(m_1) \times S_2(m_2) \\
      &&  + R_{S_1, B_2} \times S_1(m_1) \times B_2(m_2) 
        + R_{B_1, B_2} \times B_1(m_1) \times B_2(m_2) \,, \nonumber
\end{eqnarray}
where $N$ and $R$ are the corresponding normalisation parameters. 
The $\psi(2S)$ contribution is neglected in the 2-dimensional fit. 
The $\Jpsi$ pole mass and width determined from the first step are fixed in the second step of the fit, the rest of the fit parameters are left free. 
Different combinations of functional forms are used to determine the raw yield and its uncertainties. 
The signal $S$ is modelled by a Crystal Ball function including a Gaussian core and two asymmetric power-law tails~\cite{ALICE-PUBLIC-2015-006}. 
The power-law tail parameters are obtained both from data or Monte Carlo and fixed in the fits~\cite{ALICE:2017leg}. 
The background $B$ contribution is described by either the sum of two exponentials, an exponential of a second order polynomial, or the ratio of a first-order to a second-order polynomials. 
The mass distribution is fit in two different mass intervals to test the results stability, i.e. [2.0, 4.5] and [2.2, 4.9]~GeV/$c^2$.
As the candidates were assigned randomly, the fit function is symmetric under exchange of $m_1$ and $m_2$. 
The projections of one of the fits on $m_1(\mu_1^+ \mu_1^-)$ and $m_2(\mu_2^+ \mu_2^-)$ are shown in Fig.~\ref{fig:invmass}. 
The $\Jpsi$ pair signal and statistical uncertainty are evaluated as the average of the values obtained in the twelve fit configurations. The systematic uncertainty is given by their standard deviation. 
The raw yield is $N = 59.3 \pm 13.5 \, {\rm (stat)}\, \pm 4.4 \, {\rm (syst)}$. 

%
The acceptance, reconstruction and selection efficiency is evaluated assuming factorisation of the corrections of the $\Jpsi$ pair as 
\begin{equation}
    \epsilon  ({\rm J}/\psi \,\, {\rm J}/\psi )  = 
 \epsilon  ({\rm J}/\psi)  
    \times
 \epsilon  ({\rm J}/\psi)  \,.
\end{equation}
The $\Jpsi$ $\epsilon$,  $\epsilon  ({\rm J}/\psi)$, is computed from Monte Carlo simulations as described in Ref.~\cite{ALICE:2017leg}. 
An iterative procedure is used to generate input rapidity ($y$) and transverse momentum ($\pt$) distributions from data. 
The $\Jpsi$ are decayed into pairs of muons using EVTGEN~\cite{Lange:2001uf} and PHOTOS~\cite{Barberio:1993qi}. 
A GEANT3~\cite{Brun:1994aa} simulation is performed to transport the decay muons through the apparatus including a realistic description of the detector conditions during data taking. 
The validity of the factorisation approach for the efficiency calculation was tested. The invariant mass distribution was compared with the corresponding one after applying a 2-dimensional ($y
$, $\pt$) acceptance-times-efficiency correction per  $\Jpsi$ candidate. The shapes of the 2-dimensional invariant mass distribution, and their projections are not modified by the correction, confirming the validity of our assumption. 
In addition, a toy Monte Carlo was developed to study the possible influence of angular correlations between the two $\Jpsi$ of the pair. Two  $\Jpsi$ were simulated per event, according to a ($y$, $\pt$) distribution extracted from single  $\Jpsi$ measurements. To mimic possible correlations among the  $\Jpsi$, their rapidity difference was forced to follow
 either a triangular or a flat distribution. The average pair efficiency was computed for both cases. The resultant pair efficiency was found to be in agreement with the calculation from the factorisation approach for both cases.

Various sources of systematic uncertainties on the $\Jpsi$ pair production cross section are considered: 
\begin{enumerate*}[label=(\roman*)]
    \item the signal extraction, 
    \item the branching fraction uncertainty, 
    \item the luminosity normalisation, and
    \item the acceptance-times-efficiency correction. 
\end{enumerate*}
%
Details on the signal extraction procedure were given previously in this Letter. 
The systematic uncertainty on the signal extraction, obtained as described above, amounts to 7.4\%. 
%
The branching fraction uncertainty is 0.6\% for single $\Jpsi$~\cite{Workman:2022ynf}, thus 1.1\% for $\Jpsi$ pairs.
%
The influence of the luminosity normalisation factor is evaluated by computing the equivalent number of minimum-bias events in the analysed dimuon sample with different methods as described in Ref.~\cite{ALICE:2022gpu}, which amounts to 2.9\%. 
The uncertainty on the minimum bias cross section, evaluated in a van der Meer scan (1.6\%), is also taken into account in the calculation~\cite{ALICE-PUBLIC-2021-005}. 
These two sources lead to a 3.3\% systematic uncertainty for the luminosity. 
%
The systematic uncertainty on the acceptance-times-efficiency correction contains contributions from
\begin{enumerate*}[label=(\roman*)]
    \item the input $\pt$ and $y$ distributions, 
    \item the tracking efficiency in the MCH, 
    \item the MTR trigger efficiency, and
    \item the matching of the reconstructed tracks in the MCH with the track segments in the MTR.
\end{enumerate*}
The influence of the simulated $\Jpsi$ $\pt$ and $y$ distributions is tested by comparing the corrected yield obtained via the iterative procedure, with the one obtained from an efficiency-corrected invariant mass distribution.  
For this exercise, a 2-dimensional $\epsilon(\pt,y)$ correction is applied to each $\Jpsi$ candidate in order to build the efficiency-corrected invariant mass distribution, which was then fit to obtain the corresponding corrected yield. A 0.5\% uncertainty is assigned to the MC input for $\Jpsi$~\cite{ALICE:2017leg}. 
The systematic uncertainties on the tracking efficiency in the MCH, the MTR trigger efficiency and the matching between the MCH and MTR are evaluated comparing single muon data and MC, as described in Ref.~\cite{ALICE:2014uja}. The differences are then propagated to the dimuon case, being 4\%, 2\% and 1\% respectively for the $\Jpsi$~\cite{ALICE:2017leg}. 
This results in a 4.6\% acceptance-times-efficiency uncertainty for $\Jpsi$, and is propagated to a 9.2\% uncertainty for $\Jpsi$ pairs. 
The analysis requirement that all selected tracks in the MCH should match track segments in the MTR removes any possible dependence on which pair of tracks activated the trigger. 
Table~\ref{tab:unc} summarises the systematic uncertainties on the measurement of the $\Jpsi$ pair production cross section.

\begin{table}[!htbp]
    \centering
    \caption{Sources of systematic uncertainty on the $\Jpsi$ pair production cross section measurement.
    \label{tab:unc}
    }
    \begin{tabular}{l|c}
    Source                                      &  Uncertainty (\%) \\ \hline
    Signal extraction                            &  7.4  \\
    Acceptance-times-efficiency                 &  9.2  \\
    $B( {\rm J}/\psi \rightarrow \mu^+ \mu^-)$  &  1.1  \\ 
    Luminosity                                  &  3.3  \\ \hline
    Total                                       &  12.3
    \end{tabular}
\end{table}

\section{Results}
\label{sec:results}

The inclusive $\Jpsi$ pair production cross section in the kinematic interval $2.5<y<4.0$ and $\pt>0$ is measured to be 
\begin{equation}
     \sigma ({\rm J}/\psi \,\, {\rm J}/\psi ) = 10.3 \pm 2.3 \, {\rm (stat.)} \pm 1.3 \, {\rm (syst.)} \,\, {\rm nb}. \nonumber
\end{equation}
The ratio of the production cross section of the inclusive $\Jpsi$ pair to that of the inclusive $\Jpsi$ is 
\begin{equation}
     \frac{ \sigma ({\rm J}/\psi \,\, {\rm J}/\psi ) } { \sigma ({\rm J}/\psi) } = ( 9.1 \pm  2.0 \, {\rm (stat.)} \pm  1.3 \, {\rm (syst.)} ) \times 10^{-4} \,, \nonumber
\end{equation}
considering ${\rm d}\sigma ({\rm J}/\psi) / {\rm d} y = (7533.3 \pm 26.7 \, {\rm(stat.)} \pm 491.6 \, {\rm (syst.)} )$~nb for $\pt>0$ and $2.5<y<4.0$~\cite{ALICE:2017leg}, and assuming the systematic uncertainties to be uncorrelated. 
Likewise, the ratio 
\begin{equation}
    \frac{1}{2} \frac{ \sigma ({\rm J}/\psi)^2 } { \sigma ({\rm J}/\psi \,\, {\rm J}/\psi ) } = 6.2 \pm \, 1.4 \, {\rm (stat.)} \pm  1.1 \, {\rm (syst.)}  \,\, {\rm mb}, \nonumber
\end{equation}
can be calculated and interpreted as an effective cross section, according to Eq.~(\ref{eq:DPSred}). This interpretation assumes that all $\Jpsi$ pairs are produced via DPS processes. 
The relative contribution of SPS and DPS processes to $\Jpsi$ pair production is a topic of debate and intense studies, see e.g. Ref.~\cite{LHCb:2016wuo}. 
In addition, the understanding of this ratio gets challenged by the contribution of both the prompt and non-prompt components to the measured inclusive $\Jpsi$ cross section, where the non-prompt contribution originates from beauty-hadron decays.

The contamination from beauty-hadron decays to the $\Jpsi$ pair cross section is evaluated to assess the impact on the
measurement according to
\begin{equation}
    \sigma_{\rm non-prompt} ({\rm J}/\psi \,\, {\rm J}/\psi ) = \sigma_{\bbbar}^{\rm total} \times \alpha \times B^2(h_b \rightarrow \Jpsi\ \  + X)  \,.
\end{equation}
The total beauty-hadron production cross section was measured to be
\begin{displaymath}
\sigma_{\bbbar}^{\rm total} =  502 \pm 16~{\rm (stat.)} \pm 51~{\rm (syst.)} _{-3}^{+2}~{\rm (extr.)}~\mu\text{b}
\end{displaymath}
 in Ref.~\cite{ALICE:2021edd}. 
The branching ratio of a beauty hadron into a $\Jpsi$ is $B(h_b \rightarrow \Jpsi + X) = (1.16 \pm 0.10) \%$~\cite{Workman:2022ynf}, 
and the acceptance correction factor $\alpha$ is estimated using PYTHIA 8.3~\cite{Bierlich:2022pfr} simulations. 
Beauty hadrons are simulated according to three different configurations and forced to decay into $\Jpsi$. 
The three configurations use the Monash 2013 tune for the calculation~\cite{Skands:2014pea}. Two of them also include a tuning of the parameters to get a good agreement with the NLO calculation by Mangano, Nason, and Ridolfi for the $\bbbar$ single and double differential distributions~\cite{Mangano:1991jk}. The difference between the latter two is that one of them adds the ATLAS tune settings for multiple parton interactions~\cite{TheATLAScollaboration:2014rfk}. 
The $\alpha$ factor is obtained from the ratio of the $\Jpsi$ pair counts in the acceptance to the number of all $\Jpsi$ pairs in the simulation. 
The value of $\alpha = 0.044^{+0.005}_{-0.007}$ is determined as the average of the factors obtained with all configurations, and the systematic uncertainty is conservatively set to the full spread of the values. 
This gives a non-prompt contribution of
\begin{equation}
    \sigma_{\text{non-prompt}} ({\rm J}/\psi \,\, {\rm J}/\psi ) = 2.97 \pm \, 0.09 \, {\rm (stat.)} ^{+0.68}_{-0.76} {\rm (syst.)}  \,\, {\rm nb}, \nonumber
    \nonumber
\end{equation}
and correspondingly the prompt $\Jpsi$ pair cross section is
\begin{equation}
    \sigma_{\rm prompt} ({\rm J}/\psi \,\, {\rm J}/\psi )  
    =  \sigma ({\rm J}/\psi \,\, {\rm J}/\psi )  - \sigma_{\text{non-prompt}} ({\rm J}/\psi \,\, {\rm J}/\psi ) 
    = 7.3 \pm \, 1.7 \, {\rm (stat.)} ^{+1.9}_{-2.1} {\rm (syst.)}  \,\, {\rm nb}. \nonumber
\end{equation}
Analogously, for the single $\Jpsi$ case, the computed extrapolation factor to account for the number of $\Jpsi$ from beauty decays in the acceptance is $\beta=0.121 ^{+0.001}_{-0.002}$. Thus, the non-prompt contribution to the $\Jpsi$ production cross section is 
\begin{equation}
    \sigma_{\text{non-prompt}} ({\rm J}/\psi ) = 2 \times \sigma_{\bbbar}^{\rm total} \times \beta \times B(h_{\rm b} \rightarrow \Jpsi\ \  + X)  = 1.41 \pm \, 0.04 \, {\rm (stat.)} \pm 0.19 {\rm (syst.)}  \,\, \mu{\rm b}, \nonumber  
\end{equation}
and the prompt component is evaluated to be 
\begin{equation}
    \sigma_{\rm prompt} ({\rm J}/\psi ) 
    = \sigma ({\rm J}/\psi)  - \sigma_{\text{non-prompt}} ({\rm J}/\psi) 
     = 9.89 \pm \, 0.32 \, {\rm (stat.)} ^{+1.47}_{-1.48} {\rm (syst.)}  \,\, \mu{\rm b}. \nonumber
    \nonumber
\end{equation}
Therefore, the ratios discussed earlier in this section can be evaluated for the prompt case. 
The ratio of the prompt $\Jpsi$ pair production cross section to that of $\Jpsi$ equals
\begin{equation}
     \frac{ \sigma_{\rm prompt} ({\rm J}/\psi \,\, {\rm J}/\psi ) } { \sigma_{\rm prompt}  ({\rm J}/\psi) }
       = ( 7.4 \pm  1.7 \, {\rm (stat.)} \pm  2.2 \, {\rm (syst.)} ) \times 10^{-4} \,, \nonumber
\end{equation}
and the ratio related to the effective DPS cross section becomes
\begin{equation}
    \frac{1}{2} \frac{ \sigma_{\rm prompt}  ({\rm J}/\psi)^2 } { \sigma_{\rm prompt}  ({\rm J}/\psi \,\, {\rm J}/\psi ) } 
       = 6.7  \pm 1.6 \, {\rm (stat.)} \pm  2.7 \, {\rm (syst.)}  \,\, {\rm mb}. \nonumber
\end{equation}
A differential measurement of the prompt $\Jpsi$ pair production cross section and the corresponding ratios were previously reported by the LHCb collaboration in a slightly different kinematic interval, $2.0<y<4.5$ and $\pt<10$~GeV/$c$~\cite{LHCb:2011kri,LHCb:2016wuo}. The results presented here are in agreement with the LHCb ones within uncertainties.

Despite the caveat caused by the calculation of this effective value considering both the SPS and DPS contributions to the production cross section, this value is consistent with the values obtained from quarkonium-pair production measurements, with $\seff$ values ranging from 2.2 to 12.5~mb~\cite{D0:2014vql,ATLAS:2016ydt,LHCb:2016wuo,D0:2015dyx}, 
and with the values obtained for quarkonium associated production at central rapidity (in the range 2.3--6.1 mb)~\cite{Lansberg:2016muq,Lansberg:2017chq,Lansberg:2019adr}. 
It is smaller than the values obtained for associated heavy-flavour production at large rapidity by LHCb (ranging from 12.8 to 18.0 mb)~\cite{LHCb:2012aiv,LHCb:2015wvu}, or those from jet or electroweak associated production (whose values are between 12.0 and 21.3 mb)~\cite{CDF:1997yfa,D0:2009apj,CDF:1993sbj,CMS:2015wcf,CMS:2013huw,ATLAS:2013aph,CMS:2019jcb}.

\section{Conclusion}

The production cross section of $\Jpsi$ pairs at large rapidity in \pp collisions at \thirteen was studied by ALICE. The measurement exploits the full Run~2 data sample collected by ALICE. 
The production cross section of inclusive $\Jpsi$ pairs is reported to be $10.3 \pm 2.3 \, {\rm (stat.)} \pm 1.3 \, {\rm (syst.)}$~nb,  for $\Jpsi$ in the rapidity interval $2.5<y<4.0$ and for $\pt>0$. 
The effective double-parton scattering cross section is evaluated neglecting the single-parton scattering contribution. 
The results are compatible with analogous measurements performed by the LHCb collaboration in a similar kinematic interval~\cite{LHCb:2011kri,LHCb:2016wuo}.

The Run~3 data taking, with the upgraded ALICE detector and the larger accumulated luminosity~\cite{ALICE:2023udb}, will allow us to perform this measurement with increased precision and separating the prompt and non-prompt contributions. This will also enable studying the kinematics of these events and probe model calculations.


\newenvironment{acknowledgement}{\relax}{\relax}
\begin{acknowledgement}
\section*{Acknowledgements}

The ALICE Collaboration would like to thank all its engineers and technicians for their invaluable contributions to the construction of the experiment and the CERN accelerator teams for the outstanding performance of the LHC complex.
The ALICE Collaboration gratefully acknowledges the resources and support provided by all Grid centres and the Worldwide LHC Computing Grid (WLCG) collaboration.
The ALICE Collaboration acknowledges the following funding agencies for their support in building and running the ALICE detector:
A. I. Alikhanyan National Science Laboratory (Yerevan Physics Institute) Foundation (ANSL), State Committee of Science and World Federation of Scientists (WFS), Armenia;
Austrian Academy of Sciences, Austrian Science Fund (FWF): [M 2467-N36] and Nationalstiftung f\"{u}r Forschung, Technologie und Entwicklung, Austria;
Ministry of Communications and High Technologies, National Nuclear Research Center, Azerbaijan;
Conselho Nacional de Desenvolvimento Cient\'{\i}fico e Tecnol\'{o}gico (CNPq), Financiadora de Estudos e Projetos (Finep), Funda\c{c}\~{a}o de Amparo \`{a} Pesquisa do Estado de S\~{a}o Paulo (FAPESP) and Universidade Federal do Rio Grande do Sul (UFRGS), Brazil;
Bulgarian Ministry of Education and Science, within the National Roadmap for Research Infrastructures 2020¿2027 (object CERN), Bulgaria;
Ministry of Education of China (MOEC) , Ministry of Science \& Technology of China (MSTC) and National Natural Science Foundation of China (NSFC), China;
Ministry of Science and Education and Croatian Science Foundation, Croatia;
Centro de Aplicaciones Tecnol\'{o}gicas y Desarrollo Nuclear (CEADEN), Cubaenerg\'{\i}a, Cuba;
Ministry of Education, Youth and Sports of the Czech Republic, Czech Republic;
The Danish Council for Independent Research | Natural Sciences, the VILLUM FONDEN and Danish National Research Foundation (DNRF), Denmark;
Helsinki Institute of Physics (HIP), Finland;
Commissariat \`{a} l'Energie Atomique (CEA) and Institut National de Physique Nucl\'{e}aire et de Physique des Particules (IN2P3) and Centre National de la Recherche Scientifique (CNRS), France;
Bundesministerium f\"{u}r Bildung und Forschung (BMBF) and GSI Helmholtzzentrum f\"{u}r Schwerionenforschung GmbH, Germany;
General Secretariat for Research and Technology, Ministry of Education, Research and Religions, Greece;
National Research, Development and Innovation Office, Hungary;
Department of Atomic Energy Government of India (DAE), Department of Science and Technology, Government of India (DST), University Grants Commission, Government of India (UGC) and Council of Scientific and Industrial Research (CSIR), India;
National Research and Innovation Agency - BRIN, Indonesia;
Istituto Nazionale di Fisica Nucleare (INFN), Italy;
Japanese Ministry of Education, Culture, Sports, Science and Technology (MEXT) and Japan Society for the Promotion of Science (JSPS) KAKENHI, Japan;
Consejo Nacional de Ciencia (CONACYT) y Tecnolog\'{i}a, through Fondo de Cooperaci\'{o}n Internacional en Ciencia y Tecnolog\'{i}a (FONCICYT) and Direcci\'{o}n General de Asuntos del Personal Academico (DGAPA), Mexico;
Nederlandse Organisatie voor Wetenschappelijk Onderzoek (NWO), Netherlands;
The Research Council of Norway, Norway;
Commission on Science and Technology for Sustainable Development in the South (COMSATS), Pakistan;
Pontificia Universidad Cat\'{o}lica del Per\'{u}, Peru;
Ministry of Education and Science, National Science Centre and WUT ID-UB, Poland;
Korea Institute of Science and Technology Information and National Research Foundation of Korea (NRF), Republic of Korea;
Ministry of Education and Scientific Research, Institute of Atomic Physics, Ministry of Research and Innovation and Institute of Atomic Physics and University Politehnica of Bucharest, Romania;
Ministry of Education, Science, Research and Sport of the Slovak Republic, Slovakia;
National Research Foundation of South Africa, South Africa;
Swedish Research Council (VR) and Knut \& Alice Wallenberg Foundation (KAW), Sweden;
European Organization for Nuclear Research, Switzerland;
Suranaree University of Technology (SUT), National Science and Technology Development Agency (NSTDA), Thailand Science Research and Innovation (TSRI) and National Science, Research and Innovation Fund (NSRF), Thailand;
Turkish Energy, Nuclear and Mineral Research Agency (TENMAK), Turkey;
National Academy of  Sciences of Ukraine, Ukraine;
Science and Technology Facilities Council (STFC), United Kingdom;
National Science Foundation of the United States of America (NSF) and United States Department of Energy, Office of Nuclear Physics (DOE NP), United States of America.
In addition, individual groups or members have received support from:
European Research Council, Strong 2020 - Horizon 2020 (grant nos. 950692, 824093), European Union;
Academy of Finland (Center of Excellence in Quark Matter) (grant nos. 346327, 346328), Finland;
Programa de Apoyos para la Superaci\'{o}n del Personal Acad\'{e}mico, UNAM, Mexico.
\end{acknowledgement}

\bibliographystyle{utphys}   
\bibliography{alice_papers,references}

\newpage
\appendix

%
%

\section{The ALICE Collaboration}
\label{app:collab}
\begin{flushleft} 
\small
\small

S.~Acharya\,\orcidlink{0000-0002-9213-5329}\,$^{\rm 125}$, 
D.~Adamov\'{a}\,\orcidlink{0000-0002-0504-7428}\,$^{\rm 86}$, 
A.~Adler$^{\rm 69}$, 
G.~Aglieri Rinella\,\orcidlink{0000-0002-9611-3696}\,$^{\rm 32}$, 
M.~Agnello\,\orcidlink{0000-0002-0760-5075}\,$^{\rm 29}$, 
N.~Agrawal\,\orcidlink{0000-0003-0348-9836}\,$^{\rm 50}$, 
Z.~Ahammed\,\orcidlink{0000-0001-5241-7412}\,$^{\rm 132}$, 
S.~Ahmad\,\orcidlink{0000-0003-0497-5705}\,$^{\rm 15}$, 
S.U.~Ahn\,\orcidlink{0000-0001-8847-489X}\,$^{\rm 70}$, 
I.~Ahuja\,\orcidlink{0000-0002-4417-1392}\,$^{\rm 37}$
% ...

\section*{Affiliation Notes}

$^{\rm I}$ Deceased\\
$^{\rm II}$ Also at: Max-Planck-Institut f\"{u}r Physik, Munich, Germany\\
$^{\rm III}$ Also at: Italian National Agency for New Technologies, Energy and Sustainable Economic Development (ENEA), Bologna, Italy\\
$^{\rm IV}$ Also at: Department of Applied Physics, Aligarh Muslim University, Aligarh, India\\
$^{\rm V}$ Also at: Institute of Theoretical Physics, University of Wroclaw, Poland\\
$^{\rm VI}$ Also at: An institution covered by a cooperation agreement with CERN\\

\section*{Collaboration Institutes}

$^{1}$ A.I. Alikhanyan National Science Laboratory (Yerevan Physics Institute) Foundation, Yerevan, Armenia\\
$^{2}$ AGH University of Science and Technology, Cracow, Poland\\
$^{3}$ Bogolyubov Institute for Theoretical Physics, National Academy of Sciences of Ukraine, Kiev, Ukraine\\
$^{4}$ Bose Institute, Department of Physics  and Centre for Astroparticle Physics and Space Science (CAPSS), Kolkata, India\\
$^{5}$ California Polytechnic State University, San Luis Obispo, California, United States\\
$^{6}$ Central China Normal University, Wuhan, China\\
$^{7}$ Centro de Aplicaciones Tecnol\'{o}gicas y Desarrollo Nuclear (CEADEN), Havana, Cuba\\
$^{8}$ Centro de Investigaci\'{o}n y de Estudios Avanzados (CINVESTAV), Mexico City and M\'{e}rida, Mexico\\
$^{9}$ Chicago State University, Chicago, Illinois, United States\\
$^{10}$ China Institute of Atomic Energy, Beijing, China\\
$^{11}$ Chungbuk National University, Cheongju, Republic of Korea\\
$^{12}$ Comenius University Bratislava, Faculty of Mathematics, Physics and Informatics, Bratislava, Slovak Republic\\
$^{13}$ COMSATS University Islamabad, Islamabad, Pakistan\\
$^{14}$ Creighton University, Omaha, Nebraska, United States\\
$^{15}$ Department of Physics, Aligarh Muslim University, Aligarh, India\\
$^{16}$ Department of Physics, Pusan National University, Pusan, Republic of Korea\\
$^{17}$ Department of Physics, Sejong University, Seoul, Republic of Korea\\
$^{18}$ Department of Physics, University of California, Berkeley, California, United States\\
$^{19}$ Department of Physics, University of Oslo, Oslo, Norway\\
$^{20}$ Department of Physics and Technology, University of Bergen, Bergen, Norway\\
$^{21}$ Dipartimento di Fisica, Universit\`{a} di Pavia, Pavia, Italy\\
$^{22}$ Dipartimento di Fisica dell'Universit\`{a} and Sezione INFN, Cagliari, Italy\\
$^{23}$ Dipartimento di Fisica dell'Universit\`{a} and Sezione INFN, Trieste, Italy\\
$^{24}$ Dipartimento di Fisica dell'Universit\`{a} and Sezione INFN, Turin, Italy\\
$^{25}$ Dipartimento di Fisica e Astronomia dell'Universit\`{a} and Sezione INFN, Bologna, Italy\\
$^{26}$ Dipartimento di Fisica e Astronomia dell'Universit\`{a} and Sezione INFN, Catania, Italy\\
$^{27}$ Dipartimento di Fisica e Astronomia dell'Universit\`{a} and Sezione INFN, Padova, Italy\\
$^{28}$ Dipartimento di Fisica `E.R.~Caianiello' dell'Universit\`{a} and Gruppo Collegato INFN, Salerno, Italy\\
$^{29}$ Dipartimento DISAT del Politecnico and Sezione INFN, Turin, Italy\\
$^{30}$ Dipartimento di Scienze MIFT, Universit\`{a} di Messina, Messina, Italy\\
$^{31}$ Dipartimento Interateneo di Fisica `M.~Merlin' and Sezione INFN, Bari, Italy\\
$^{32}$ European Organization for Nuclear Research (CERN), Geneva, Switzerland\\
$^{33}$ Faculty of Electrical Engineering, Mechanical Engineering and Naval Architecture, University of Split, Split, Croatia\\
$^{34}$ Faculty of Engineering and Science, Western Norway University of Applied Sciences, Bergen, Norway\\
$^{35}$ Faculty of Nuclear Sciences and Physical Engineering, Czech Technical University in Prague, Prague, Czech Republic\\
$^{36}$ Faculty of Physics, Sofia University, Sofia, Bulgaria\\
$^{37}$ Faculty of Science, P.J.~\v{S}af\'{a}rik University, Ko\v{s}ice, Slovak Republic\\
$^{38}$ Frankfurt Institute for Advanced Studies, Johann Wolfgang Goethe-Universit\"{a}t Frankfurt, Frankfurt, Germany\\
$^{39}$ Fudan University, Shanghai, China\\
$^{40}$ Gangneung-Wonju National University, Gangneung, Republic of Korea\\
$^{41}$ Gauhati University, Department of Physics, Guwahati, India\\
$^{42}$ Helmholtz-Institut f\"{u}r Strahlen- und Kernphysik, Rheinische Friedrich-Wilhelms-Universit\"{a}t Bonn, Bonn, Germany\\
$^{43}$ Helsinki Institute of Physics (HIP), Helsinki, Finland\\
$^{44}$ High Energy Physics Group,  Universidad Aut\'{o}noma de Puebla, Puebla, Mexico\\
$^{45}$ Horia Hulubei National Institute of Physics and Nuclear Engineering, Bucharest, Romania\\
$^{46}$ Indian Institute of Technology Bombay (IIT), Mumbai, India\\
$^{47}$ Indian Institute of Technology Indore, Indore, India\\
$^{48}$ INFN, Laboratori Nazionali di Frascati, Frascati, Italy\\
$^{49}$ INFN, Sezione di Bari, Bari, Italy\\
$^{50}$ INFN, Sezione di Bologna, Bologna, Italy\\
$^{51}$ INFN, Sezione di Cagliari, Cagliari, Italy\\
$^{52}$ INFN, Sezione di Catania, Catania, Italy\\
$^{53}$ INFN, Sezione di Padova, Padova, Italy\\
$^{54}$ INFN, Sezione di Pavia, Pavia, Italy\\
$^{55}$ INFN, Sezione di Torino, Turin, Italy\\
$^{56}$ INFN, Sezione di Trieste, Trieste, Italy\\
$^{57}$ Inha University, Incheon, Republic of Korea\\
$^{58}$ Institute for Gravitational and Subatomic Physics (GRASP), Utrecht University/Nikhef, Utrecht, Netherlands\\
$^{59}$ Institute of Experimental Physics, Slovak Academy of Sciences, Ko\v{s}ice, Slovak Republic\\
$^{60}$ Institute of Physics, Homi Bhabha National Institute, Bhubaneswar, India\\
$^{61}$ Institute of Physics of the Czech Academy of Sciences, Prague, Czech Republic\\
$^{62}$ Institute of Space Science (ISS), Bucharest, Romania\\
$^{63}$ Institut f\"{u}r Kernphysik, Johann Wolfgang Goethe-Universit\"{a}t Frankfurt, Frankfurt, Germany\\
$^{64}$ Instituto de Ciencias Nucleares, Universidad Nacional Aut\'{o}noma de M\'{e}xico, Mexico City, Mexico\\
$^{65}$ Instituto de F\'{i}sica, Universidade Federal do Rio Grande do Sul (UFRGS), Porto Alegre, Brazil\\
$^{66}$ Instituto de F\'{\i}sica, Universidad Nacional Aut\'{o}noma de M\'{e}xico, Mexico City, Mexico\\
$^{67}$ iThemba LABS, National Research Foundation, Somerset West, South Africa\\
$^{68}$ Jeonbuk National University, Jeonju, Republic of Korea\\
$^{69}$ Johann-Wolfgang-Goethe Universit\"{a}t Frankfurt Institut f\"{u}r Informatik, Fachbereich Informatik und Mathematik, Frankfurt, Germany\\
$^{70}$ Korea Institute of Science and Technology Information, Daejeon, Republic of Korea\\
$^{71}$ KTO Karatay University, Konya, Turkey\\
$^{72}$ Laboratoire de Physique des 2 Infinis, Ir\`{e}ne Joliot-Curie, Orsay, France\\
$^{73}$ Laboratoire de Physique Subatomique et de Cosmologie, Universit\'{e} Grenoble-Alpes, CNRS-IN2P3, Grenoble, France\\
$^{74}$ Lawrence Berkeley National Laboratory, Berkeley, California, United States\\
$^{75}$ Lund University Department of Physics, Division of Particle Physics, Lund, Sweden\\
$^{76}$ Nagasaki Institute of Applied Science, Nagasaki, Japan\\
$^{77}$ Nara Women{'}s University (NWU), Nara, Japan\\
$^{78}$ National and Kapodistrian University of Athens, School of Science, Department of Physics , Athens, Greece\\
$^{79}$ National Centre for Nuclear Research, Warsaw, Poland\\
$^{80}$ National Institute of Science Education and Research, Homi Bhabha National Institute, Jatni, India\\
$^{81}$ National Nuclear Research Center, Baku, Azerbaijan\\
$^{82}$ National Research and Innovation Agency - BRIN, Jakarta, Indonesia\\
$^{83}$ Niels Bohr Institute, University of Copenhagen, Copenhagen, Denmark\\
$^{84}$ Nikhef, National institute for subatomic physics, Amsterdam, Netherlands\\
$^{85}$ Nuclear Physics Group, STFC Daresbury Laboratory, Daresbury, United Kingdom\\
$^{86}$ Nuclear Physics Institute of the Czech Academy of Sciences, Husinec-\v{R}e\v{z}, Czech Republic\\
$^{87}$ Oak Ridge National Laboratory, Oak Ridge, Tennessee, United States\\
$^{88}$ Ohio State University, Columbus, Ohio, United States\\
$^{89}$ Physics department, Faculty of science, University of Zagreb, Zagreb, Croatia\\
$^{90}$ Physics Department, Panjab University, Chandigarh, India\\
$^{91}$ Physics Department, University of Jammu, Jammu, India\\
$^{92}$ Physics Program and International Institute for Sustainability with Knotted Chiral Meta Matter (SKCM2), Hiroshima University, Hiroshima, Japan\\
$^{93}$ Physikalisches Institut, Eberhard-Karls-Universit\"{a}t T\"{u}bingen, T\"{u}bingen, Germany\\
$^{94}$ Physikalisches Institut, Ruprecht-Karls-Universit\"{a}t Heidelberg, Heidelberg, Germany\\
$^{95}$ Physik Department, Technische Universit\"{a}t M\"{u}nchen, Munich, Germany\\
$^{96}$ Politecnico di Bari and Sezione INFN, Bari, Italy\\
$^{97}$ Research Division and ExtreMe Matter Institute EMMI, GSI Helmholtzzentrum f\"ur Schwerionenforschung GmbH, Darmstadt, Germany\\
$^{98}$ Saga University, Saga, Japan\\
$^{99}$ Saha Institute of Nuclear Physics, Homi Bhabha National Institute, Kolkata, India\\
$^{100}$ School of Physics and Astronomy, University of Birmingham, Birmingham, United Kingdom\\
$^{101}$ Secci\'{o}n F\'{\i}sica, Departamento de Ciencias, Pontificia Universidad Cat\'{o}lica del Per\'{u}, Lima, Peru\\
$^{102}$ Stefan Meyer Institut f\"{u}r Subatomare Physik (SMI), Vienna, Austria\\
$^{103}$ SUBATECH, IMT Atlantique, Nantes Universit\'{e}, CNRS-IN2P3, Nantes, France\\
$^{104}$ Sungkyunkwan University, Suwon City, Republic of Korea\\
$^{105}$ Suranaree University of Technology, Nakhon Ratchasima, Thailand\\
$^{106}$ Technical University of Ko\v{s}ice, Ko\v{s}ice, Slovak Republic\\
$^{107}$ The Henryk Niewodniczanski Institute of Nuclear Physics, Polish Academy of Sciences, Cracow, Poland\\
$^{108}$ The University of Texas at Austin, Austin, Texas, United States\\
$^{109}$ Universidad Aut\'{o}noma de Sinaloa, Culiac\'{a}n, Mexico\\
$^{110}$ Universidade de S\~{a}o Paulo (USP), S\~{a}o Paulo, Brazil\\
$^{111}$ Universidade Estadual de Campinas (UNICAMP), Campinas, Brazil\\
$^{112}$ Universidade Federal do ABC, Santo Andre, Brazil\\
$^{113}$ University of Cape Town, Cape Town, South Africa\\
$^{114}$ University of Houston, Houston, Texas, United States\\
$^{115}$ University of Jyv\"{a}skyl\"{a}, Jyv\"{a}skyl\"{a}, Finland\\
$^{116}$ University of Kansas, Lawrence, Kansas, United States\\
$^{117}$ University of Liverpool, Liverpool, United Kingdom\\
$^{118}$ University of Science and Technology of China, Hefei, China\\
$^{119}$ University of South-Eastern Norway, Kongsberg, Norway\\
$^{120}$ University of Tennessee, Knoxville, Tennessee, United States\\
$^{121}$ University of the Witwatersrand, Johannesburg, South Africa\\
$^{122}$ University of Tokyo, Tokyo, Japan\\
$^{123}$ University of Tsukuba, Tsukuba, Japan\\
$^{124}$ University Politehnica of Bucharest, Bucharest, Romania\\
$^{125}$ Universit\'{e} Clermont Auvergne, CNRS/IN2P3, LPC, Clermont-Ferrand, France\\
$^{126}$ Universit\'{e} de Lyon, CNRS/IN2P3, Institut de Physique des 2 Infinis de Lyon, Lyon, France\\
$^{127}$ Universit\'{e} de Strasbourg, CNRS, IPHC UMR 7178, F-67000 Strasbourg, France, Strasbourg, France\\
$^{128}$ Universit\'{e} Paris-Saclay Centre d'Etudes de Saclay (CEA), IRFU, D\'{e}partment de Physique Nucl\'{e}aire (DPhN), Saclay, France\\
$^{129}$ Universit\`{a} degli Studi di Foggia, Foggia, Italy\\
$^{130}$ Universit\`{a} del Piemonte Orientale, Vercelli, Italy\\
$^{131}$ Universit\`{a} di Brescia, Brescia, Italy\\
$^{132}$ Variable Energy Cyclotron Centre, Homi Bhabha National Institute, Kolkata, India\\
$^{133}$ Warsaw University of Technology, Warsaw, Poland\\
$^{134}$ Wayne State University, Detroit, Michigan, United States\\
$^{135}$ Westf\"{a}lische Wilhelms-Universit\"{a}t M\"{u}nster, Institut f\"{u}r Kernphysik, M\"{u}nster, Germany\\
$^{136}$ Wigner Research Centre for Physics, Budapest, Hungary\\
$^{137}$ Yale University, New Haven, Connecticut, United States\\
$^{138}$ Yonsei University, Seoul, Republic of Korea\\
$^{139}$  Zentrum  f\"{u}r Technologie und Transfer (ZTT), Worms, Germany\\
$^{140}$ Affiliated with an institute covered by a cooperation agreement with CERN\\
$^{141}$ Affiliated with an international laboratory covered by a cooperation agreement with CERN.\\

\end{flushleft} 
  
\end{document}